# Coupling of optical phonons with Kondo effect and magnetic orders in antiferromagnetic Kondo lattice $CeAuSb_2$

Yin-Zou Zhao,[1] Qi-Yi Wu,[1] Chen Zhang,[1] Bo Chen,[1] Wei Xia,[2] Jiao-Jiao Song,[1] Ya-Hua Yuan,[1] Hao Liu,[1] Fan-Ying Wu,[1] Xue-Qing Ye,[1] Hong-Yi Zhang,[1] Han Huang,[1] Hai-Yun Liu,[3] Yu-Xia Duan,[1] Yan-Feng Guo,[2,4] Jun He,[1] and Jian-Qiao Meng[1,*]

[1]*School of Physics, Central South University, Changsha 410083, Hunan, China*
[2]*School of Physical Science and Technology, ShanghaiTech University, Shanghai 201210, China*
[3]*Beijing Academy of Quantum Information Sciences, Beijing 100085, China*
[4]*ShanghaiTech Laboratory for Topological Physics, ShanghaiTech University, Shanghai 201210, China*

Ultrafast optical spectroscopy was used to investigate the ultrafast quasiparticle dynamics of antiferromagnetic Kondo lattice $CeAuSb_2$ as a function of temperature and fluence. Our results reveal (i) the opening of a narrow hybridization gap ($\Delta \sim 4.5$ meV) near the Fermi level below the coherence temperature $T^* \approx 100$ K, (ii) the coupling of coherent phonon modes with Kondo effect and magnetic orders, leading to the frequencies anomaly at the characteristic temperatures, and (iii) a possible photoinduced nonthermal phase transition. Our observations thus shed light on the hybridization dynamics and magnetic orders in heavy fermion systems.

The coupling between spin, orbital, charge, and lattice degrees of freedom in strongly correlated electronic systems leads to rich phase diagrams. For heavy fermion (HF) materials with partially filled $4f$ or $5f$ orbitals, the competition between Kondo and Ruderman-Kittel-Kasuya-Yosida (RKKY) interaction dominates, leading to the existence of complex magnetic ground states [1–3]. The $f$ electrons are initially localized at high temperatures and gradually become partially itinerant as the temperature decreases due to hybridization with conducting electrons, characterizing the duality of localization and itinerancy at low temperatures [4–8]. Phonons are considered to play an essential role in some particular HF systems, such as $CeCu_2Si_2$ [9], $CeCoIn_5$ [10], $USb_2$ [11], and $Yb_{14}MnSb_{11}$ [12]. It has been suggested that coupling phonon with the Kondo effect or magnetism may lead to new properties [13–15], such as superconductivity [13, 14]. However, very few works have been done on this issue, partly because it is challenging to detect such a coupling. The stoichiometric $CeAuSb_2$, with well-balanced Kondo and RKKY interactions, exhibits a complex low-temperature magnetic phase diagram [16–20], making it a good platform for studying this concern.

Layered $CeAuSb_2$ is a member of $CeTX_2$ ($T$ = Au, Ag, Ni, Cu, Pd; $X$ = Sb, Bi), a family of HF compounds with pronounced crystalline electric field (CEF) effects [20–26]. $CeAuSb_2$ crystallizes in the tetragonal crystal structure ($P4/nmm$) [21]. The resistivity measurement revealed that the Kondo effect [17–19], or the onset of short-range magnetic order [17], or a combination of the two [17], occurs at around 14 K. At ambient pressure, it undergoes an antiferromagnetic (AFM) phase transition at the Néel temperature $T_N$ = 5-6.8 K, depending on the occupancy of the Au site [19–21]. Transport measurements suggest a large magnetic anisotropy with an AFM easy axis along the [001] direction [21]. $T_N$ can be continuously suppressed with magnetic fields applied along the [001] direction, leading to a quantum critical point (QCP) at $H_c \simeq 5.4$ T [19–21]. At a temperature slightly above $T_N$ = 6.3 K, bulk thermodynamic probes and x-ray diffraction reveal a nematic transition at $T_{nem}$ = 6.5 K [27], which is widely observed in iron-based superconductors and is considered to be intertwined with superconductivity [28]. $CeAuSb_2$ may also manifest multiple magnetic phase transitions at lower temperatures [29].

Ultrafast optical spectroscopy is an effective method for studying the complex behavior of low-energy electrons. It offers insights into the physics of correlated materials [30], such as transition metal dichalcogenides [31–33], high-temperature superconductors [34–37], and HFs [10–12, 38–41]. By conducting time-resolved out-of-equilibrium studies, different degrees of freedom (quasiparticles) can be disentangled in the temporal domain based on their varying lifetimes. In particular, ultrafast spectroscopy is very sensitive to changes in the low-energy electronic structure, such as the opening of a narrow energy gap near the Fermi energy ($E_F$) due to the slow relaxation process caused by the bottleneck effect [10–12, 35–41].

Here in this paper, we report an ultrafast optical spectroscopy study of the response of $CeAuSb_2$ single crystal after it has been driven out of equilibrium with an ultrashort laser pulse. The temperature dependence of the quasiparticle relaxation reveals a coherence temperature $T^* \approx 100$ K caused by the onset of a narrow hybridization gap ($\Delta \sim 4.5$ meV). Measurements under high pump fluence reveal two coherent phonon modes with $\sim 3.66$ and 3.48 THz frequencies ($T$ = 3.8 K), respectively.The amplitudes, relaxation times, and phonon frequencies show anomalous behavior at around $T^*$, $T'$($\approx 13$ K), and $T_N$/$T_{nem}$. We believe the Kondo effect and magnetic order have essential effects on the quasiparticle dynamics. We also reveal a possible photoinduced nonthermal phase transition by pumping fluence-dependent measurements.

High-quality single crystal $CeAuSb_2$ was grown from Sb flux. The temperature dependence of the in-plane resistivity $\rho$ shows a sharp drop around 7 K, with a broad shoulder around $T' \approx 13$ K (see Fig. S1 in the Supplemental Material [42]). The former corresponds to the AFM transition temperature $T_N$, while the latter is more complex and will be discussed later. Ultrafast differential reflectivity $\Delta R/R$ was performed at a center wavelength of 800 nm ($\sim 1.55$ eV), using a 1 MHz Yb-based femtosecond laser oscillator with a pulse width of $\sim$

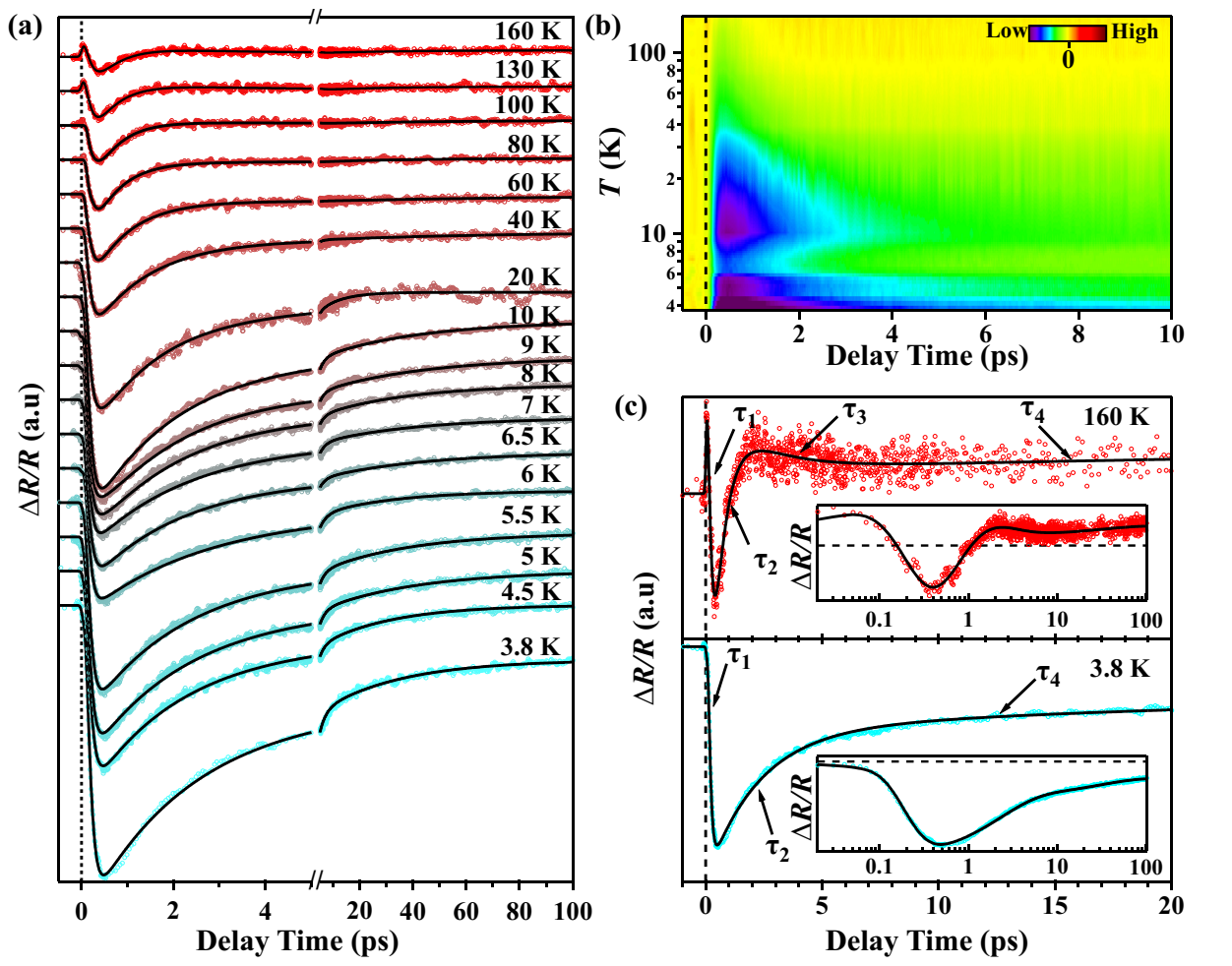

FIG. 1. (color online)(**a**) $\Delta R/R$ as a function of delay time over a temperature range from 3.8 to 160 K at pump fluence of 30 $\mu$J/cm$^2$. The black solid lines are Eq. (1) fits. Note the break in the $x$ axis. (**b**) 2D pseudocolor map of $\Delta R/R$ as a function of temperature and delay time. **(c)** Illustration of fitting using Eq. (1) (upper panel:160 K; lower panel: 3.8 K). The arrows indicate the corresponding relaxation processes. The insets show the dynamics at long time scales. For 3.8 K, $A_3$ was set to zero.

35 fs. The pump and probe beam spot diameters on the sample were $\sim$ 140 $\mu$m and $\sim$ 60 $\mu$m, respectively. The pump and probe beams were $s$- and $p$-polarized, respectively. All measurements were taken in a high vacuum (better than $10^{-6}$ Torr). The details of the experimental setup are described elsewhere [36].

Figure 1(a) presents differential reflectivity $\Delta R/R$ as a function of delay time over a temperature range from 3.8 to 160 K at a low pump fluence of 30 $\mu$J/cm$^2$. Photoexcitation results in instantaneous changes in $\Delta R/R$ due to an increase in the temperature of the electronic system, followed by multiple recovery processes. The recovery processes are dominated by the electron-electron ($e$-$e$) and electron-boson scattering processes. The transient reflectivity shows strong temperature dependence. As the temperature falls, the initial rapid rise weakens, and only a tiny peak remains at low temperatures (see Fig. S2 in the Supplemental Material [42] for the initial rise); the hump that appears at about 2 ps due to the second rising process also diminishes, disappearing at about 100 K. The second rising process observed at high temperatures is likely related to the coupling of quasiparticles with some bosonic excitations, which is widely present in strongly correlated materials [10, 11, 36–38, 43], semimetals [32], and topological insulators [44]. The final relatively slow relaxation at all temperatures may be due to thermal diffusion, as reported for other strongly corrected materials [11, 41].

Figure 1(b) is a 2D pseudocolor mapping image of $\Delta R/R$ as a function of pump-probe delay time ($x$ axis) and temperature ($y$ axis). To highlight the drastic changes in transient reflectivity at low temperatures and at the beginning, the temperature axis is plotted in logarithmic coordinates and the time axis is given only in the range of a few ps. It can be seen that $\Delta R/R$ shows noticeable changes around $T_N/T_{nem}$ and 10-

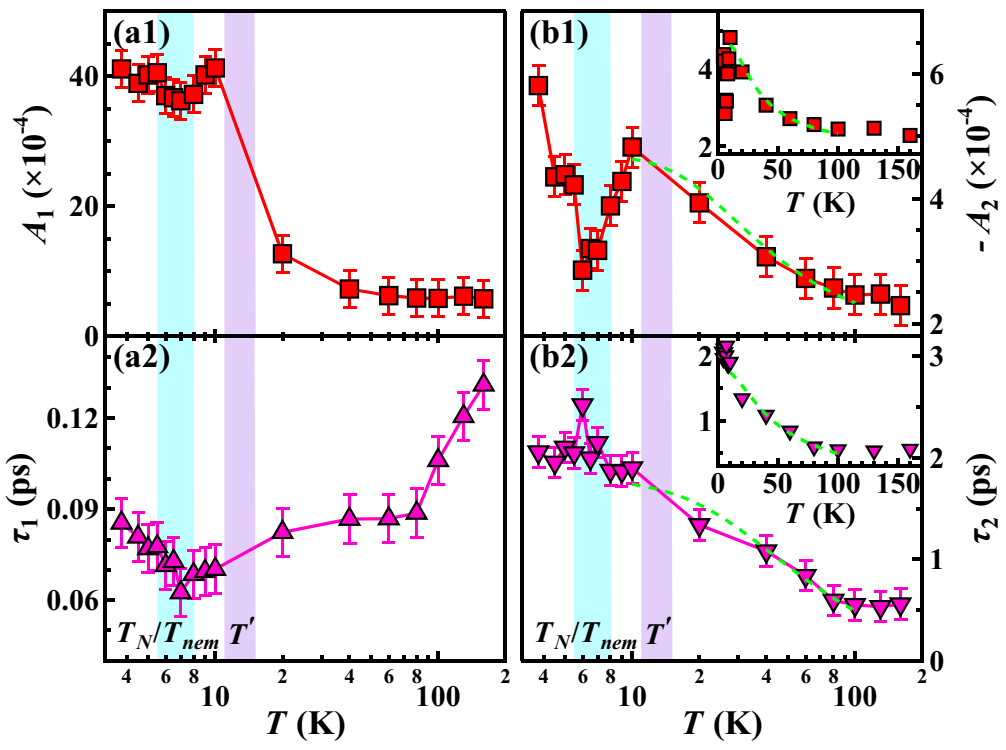

FIG. 2. (color online) **(a1)**, **(a2)** Extracted amplitude $A_1$ and lifetime $\tau_1$ as a function of temperature, respectively. **(b1)**, **(b2)** Extracted amplitude $A_2$ and lifetime $\tau_2$ as a function of temperature, respectively. The dashed green lines show the RT model fitting curves.

20 K. A quantitative analysis of quasiparticle dynamics was conducted to investigate its temperature-dependent behavior. The solid black lines in Figs. 1(a) and 1(c) suggest that the transient reflectivity over a considerable time domain (up to 100 ps) fits well with

$$\frac{\Delta R(t)}{R} = \frac{1}{\sqrt{2\pi}w}\exp(-\frac{t^2}{2w^2})\otimes[\sum_{i=1}^{4}A_i\exp(-\frac{t-t_0}{\tau_i})]+C \quad (1)$$

where $A_i$ and $\tau_i$ are the amplitude and relaxation time of the $i$th decay process, respectively. $w$ is the incidence pulse temporal duration and $C$ is a constant offset. At higher $T$, the entire recovery time window is better fit by a quad-exponential decay; the triexponential fit function can account for the data well for $T \leq 100$ K ($A_3$ was set to zero).

The extracted amplitudes and relaxation times of the first and second components are shown in Fig. 2 as a function of temperature. The first and briefest relaxation process, with a lifetime $\tau_1$ of about 0.1 ps, is usually considered an $e$-$e$ scattering process [Fig. 2(a2)] [36]. Two evident anomalies at low temperatures are found. One anomaly occurs around 7 K, roughly equal to $T_N/T_{nem}$, indicating the AFM or nematic order or both are operating. The other occurs between 10 and 20 K [Fig. 2(a1)] , indicating some transition had occurred. Our (see Fig. S1 in the Supplemental Material [42]) and previous resistivity measurements have observed anomalies around this temperature, $T' \approx 13$ K [17–19]. This is thought to be associated with the Kondo effect [17–19], or the onset of short-range magnetic order [17], or a combination of the two [17]. The underlying mechanism of this anomaly is discussed in more detail later.

As shown in the inset of Figs. 2(b1) and 2(b2), both -$A_2$ and relaxation time $\tau_2$ increase with decreasing temperature below $T^* \approx 100$ K. Here, $T^*$ is close to the first excitation CEF level $|\pm\frac{3}{2}\rangle$ (97 K) [21] and roughly corresponds to the transport anomaly caused by the thermal occupation of the excited CEF state [17], indicating that the heavy electron coherence and relaxation are closely related. A similar phenomenon that the lifetime or amplitude increases suddenly with the decrease of temperature was commonly reported in other strongly correlated materials, such as heavy fermions [10, 38] and high-temperature superconductors [36, 37], and the anomaly temperature is considered to be the onset of hybridization gap and pseudogap opening, respectively. The temperature dependence of decay rate and amplitude can be well described by the Rothwarf-Taylor (RT) model [45], which successfully describes the correlated system with a narrow energy gap near $E_F$ [10–12, 35–41]. According to the RT model, the gap function can be quantitatively extracted from the temperature-dependent $\tau(T)$ and $A(T)$ [37, 40, 41],

$$A(T) \propto \frac{\varepsilon_I/[\Delta + k_B T/2]}{1+\gamma\sqrt{2k_B T/\pi\Delta}e^{-\Delta/k_B T}}, \quad (2)$$

$$\tau(T) \propto \frac{1}{[\delta(\beta n_T+1)^{-1}+2n_T](\Delta+\alpha T\Delta^4)} \quad (3)$$

$$n_T(T) = \frac{A(0)}{A(T)} - 1 \propto (T\Delta)^p e^{-\Delta/T} \quad (4)$$

where $\varepsilon_I$ is the absorbed laser energy density per unit cell, $n_T$ is the density of quasiparticles thermally excited across the gap, and $\alpha$, $\beta$, $\gamma$, and $\delta$ are fitting parameters. The value of $p$ $(0 < p < 1)$ depends on the shape of the gapped density of state (DOS). As shown in Figs. 2(b1) and 2(b2), a good fit was achieved below $T^*$ using a $T$-independent constant gap $\Delta$, yielding a value of 4.5 meV with $p = 0.5$ from the Bardeen-Cooper-Schrieffer (BCS) superconductor form of the DOS [41]. Below $T'$, the decrease of $A_2$ and the saturation of $\tau_2$ suggest that the hybridization process may be suppressed. In Ce-based HF materials, the Kondo effect and magnetic order promote and inhibit hybridization, respectively [46, 47]. This means that the transition occurring at $T'$ is unlikely to be associated with the Kondo effect [17–19], but rather the onset of short-range magnetic order [17] or a combination of the Kondo effect and the onset of short-range magnetic order [17].

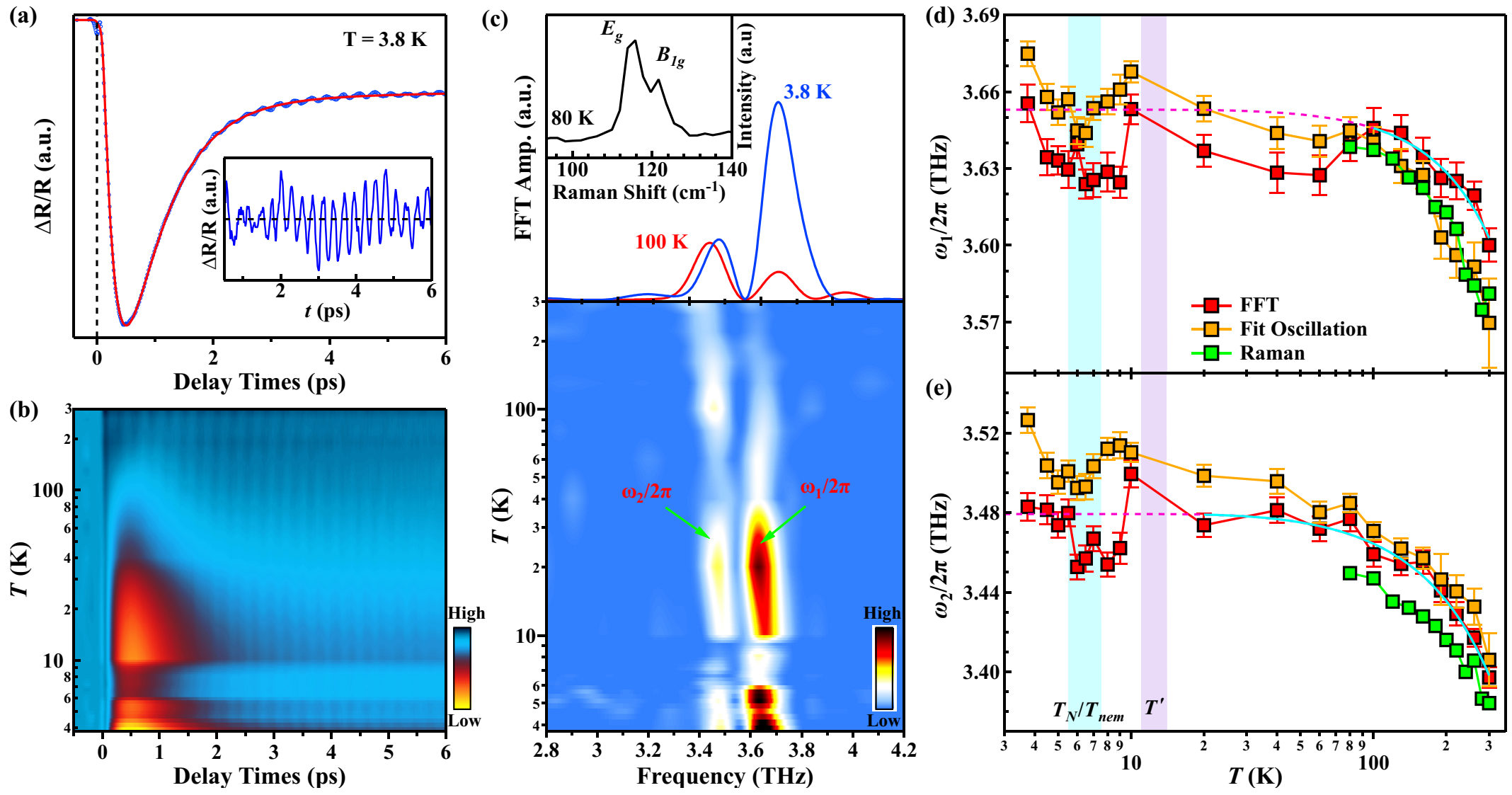

FIG. 3. (color online) (**a**) Transient reflectivity $\Delta R/R$ is shown as a function of delay time at 3.8 K at a pump fluence of 150 $\mu$J/cm$^2$, showing coherent optical phonon vibration superimposed on quasiparticle relaxation. The solid line is Eq. (1) fit. Inset: oscillations after subtracting nonoscillatory background. (**b**) 2D pseudocolor map of $\Delta R/R$ as a function of delay time over a temperature range from 3.8 to 300 K. (**c**) FFT spectrum in the frequency domain for the extracted oscillations from 3.8 up to 300 K. Two coherent phonon modes, $\omega_1$ and $\omega_2$, can be seen. Inset: Raman spectrum at 80 K. (**d**), (**e**) The derived frequencies of $\omega_1$ and $\omega_2$ as a function of temperature. The solid cyan lines are the fitted curves of the FFT data (red squares) using an anharmonic phonon model. And the magenta dashed lines are the extension of the anharmonic fitting.

Having explored the transient reflectivity at low pump fluence, we now study the collective bosonic excitation, phonon, which is a good measure of the phase transition [10, 12, 38]. At a high pump fluence of 150 $\mu$J/cm$^2$, the temperature evolution of phonon frequencies was investigated to reveal the internal relationship of phonons with Kondo interaction and complex magnetic orders. Significant high-frequency oscillations occurred instantaneously upon photoexcitation and superimposed on the $\Delta R/R$ profile [Fig. 3(a)]. Oscillations and quasiparticle dynamics contributed to the initial reflectivity peak (see Fig. S3 in the Supplemental Material [42]). Figure 3(b) is a 2D pseudocolor $\Delta R/R$ mapping shown as a function of pump-probe delay and temperature. Oscillations can be observed at all measured temperatures (see Fig. S3 in the Supplemental Material [42]). The time-domain oscillation, extracted by subtracting nonoscillatory decay, was shown in the inset of Fig. 3(a).

Figure 3(c) shows the oscillatory components extracted by fast Fourier transform (FFT) on the oscillations after subtracting the nonoscillatory background. Two distinct terahertz modes were observed at all measured temperatures, including $\omega_1/2\pi \sim 3.66$ THz (i.e., 15.1 meV or 122.1 cm$^{-1}$) and $\omega_2/2\pi \sim 3.48$ THz (i.e., 14.4 meV or 116.1 cm$^{-1}$) at 3.8 K. Both $\omega_1$ and $\omega_2$ frequencies vary significantly with temperature. The inset of Fig. 3(c) shows the Raman spectrum of $CeAuSb_2$ single crystal, which was obtained in the backscattering configuration under 633 nm laser excitation at 80 K. Two phonon modes are observed within the display wavelength range (see

Figs. S5 in the Supplemental Material [42] for more Raman spectra). The phonon modes at $\sim$121.4 cm$^{-1}$ (i.e., 15.1 meV or 3.64 THz) and $\sim$ 115.1 cm$^{-1}$ (i.e., 14.3 meV or 3.45 THz) can be assigned to $B_{1g}$ and $E_g$ modes, respectively, based on the comparison of Raman spectra (see Fig. S5 in the Supplemental Material [42]) with the calculated phonon dispersion(see Fig. S4 and Table 1 in the Supplemental Material [42]). $\omega_1$ and $\omega_2$ are identical to $B_{1g}$ and $E_g$ phonon modes, respectively. This suggests that $\omega_1$ and $\omega_2$ can be identified as the coherent $B_{1g}$ and $E_g$ phonon modes of Sb atoms, respectively.

The oscillation components also can be fitted quantitatively using the expression

$$\left(\frac{\Delta R(t)}{R}\right)_{osc} = \sum_{j=1,2} A_i e^{-\Gamma_j t} \sin(\omega_j t + \varphi_j) \tag{5}$$

where $A_j$, $\Gamma_j$, $\omega_j$, and $\varphi_j$ are the amplitude, damping rate, frequency, and phase, respectively. The extracted temperature dependence of $\omega_1$ and $\omega_2$ frequencies are plotted in Figs. 3(d) and 3(e), respectively. $B_{1g}$ and $E_g$ frequencies were estimated by fitting the Raman spectra peak with a Fano resonance line shape [48, 49]. $\omega_1$ ($B_{1g}$) and $\omega_2$ ($E_g$) harden with decreasing temperature. $\omega_1$ and $\omega_2$ can be well fitted by the optical phonon anharmonic effects (cyan lines) [50, 51] above $T^*$ and $T'$, respectively. Below $T^*$, $\omega_1$ shows a weak renormalization, leading to a slight deviation from the monotonic increase with lowering temperature (magenta dashed line). The anomalous softening of $\omega_1$ mode and the opening of the hybridization gap occur at the same temperature $T^*$, indicating the coupling between $\omega_1$ mode and Kondo interaction, consistent with the results reported in other HF compounds [10–12, 38]. In contrast, $\omega_2$ does not seem to be affected by the opening of the hybridization gap. In addition, $\omega_1$ and $\omega_2$ show anomalies across $T'$ and $T_N/T_{nem}$. It has been suggested that *e-ph* coupling is a possible origin of the electronic nematic phase in the iron-based superconductor $BaFe_2As_2$ [52] and topological superconductor candidate $Sr_{0.1}Bi_2Se_2$ [53]. These observations suggest that *e-ph* coupling is closely related to the DOS change near $E_F$ in the presence of hybridization [10, 38] and is also associated with magnetic orders [52, 53]. Our results imply that the magnon-phonon coupling may be involved in the low-energy band renormalization [44].

Near the quantum critical point, a magnetic transition can be suppressed to zero temperature by tuning the ratio of Kondo and RKKY interactions, which is determined by nonthermal parameters such as pressure, magnetic field, or chemical doping. Additionally, nonthermal phase transitions can also be induced by intense laser fluence. $CeAuSb_2$ is an ideal platform to study possible photoinduced phase transition due to its low magnetic transition temperature and well-balanced Kondo and RKKY interactions.

Fluence-dependent measurements were performed at 3.8 K to investigate possible photoinduced phase transition, as shown in Fig. 4(a). It can be seen that the $\Delta R/R$ signal varies significantly with the pump fluence. With the increase of fluence, a strong initial peak right after pumping and a second rise peak are formed. Similar fluence-dependent behavior has been observed in the ferromagnetic Kondo lattice compound $CeAgSb_2$ [54]. It should be noted that although there are some similarities between the high fluence data [Fig. 4(a)] and the high-temperature data [Fig. 1(a); also see Figs. S2 and S3 in the Supplemental Material [42]], there are significant differences in details. The inset of Fig. 4(a) displays the fluence dependence of the location of the reflectivity signal minimum $t_{min}$. As the fluence increases, $t_{min}$ is a constant below the threshold $F_c$ ($\sim$ 200 $\mu$J/cm$^2$). It increases steadily from $\sim$ 0.5 ps to $\sim$ 1.4 ps above $F_c$, marking the onset of different relaxation processes of quasiparticles. To explain such a fluence threshold $F_C$, we consider three possibilities: (i) intense pumping causes a significant heating effect, leading to an increase in the temperature of the illuminated part of the sample, particularly at low temperatures; (ii) intense pumping that excites enough quasiparticles near the $E_F$ to the higher energy levels, leading to the smearing of the hybridization gap [41]; (iii) photoinduced quenching of the magnetic order [55]. The latter two possibilities may also be two aspects of the same phase transition, i.e., photoinduced magnetic order quenching accompanied by hybridization gap smearing.

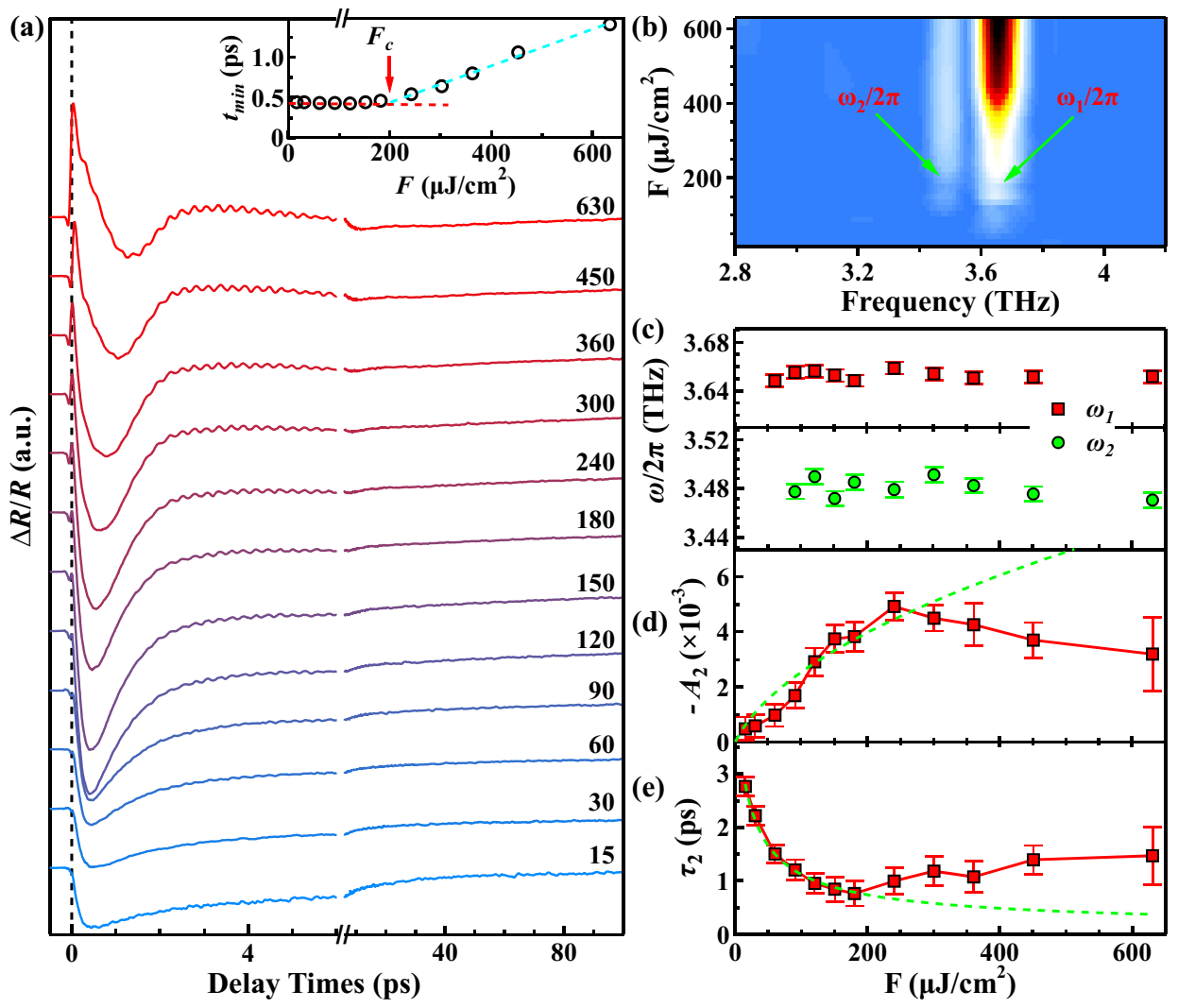

FIG. 4. (color online)(**a**) Fluence dependence of the $\Delta R/R$ as a function of delay time at 3.8 K. Note the break in the $x$ axis. Inset: fluence dependence of the location of the reflectivity minimum. The dashed red and cyan lines serve as a guide for the eyes. (**b**) FFT spectrum color intensity map as a function of frequency and fluences. (**c**) Derived frequencies of $\omega_1$ and $\omega_2$ as a function of pump fluence. (**d**) Fluence dependence of -$A_2$, together with the fit (dashed green line), given by Eq. (6). (**e**) Fluence dependence of $\tau_2$. The dashed green line is the fit to the data with RT model.

Figure 4(b) displays the FFT spectrum as a function of frequency and pump fluence. $\omega_1$ and $\omega_2$ frequency variances are petty. Figure 4(c) shows the frequencies of $\omega_1$ and $\omega_2$ extracted by an FFT method as a function of pump fluence. Typically, phonons soften with increasing fluence at high pump fluence due to a significant heating effect [32, 36]. However, the frequencies of $\omega_1$ and $\omega_2$ almost do not vary with pump fluence, and thus significant thermal effects can be excluded since $\omega_1$ and $\omega_2$ show a strong temperature dependence at a pump fluence of 150 $\mu$J/cm$^2$ [Figs. 3(d) and 3(e)]. What is the reason for the anomalous fluence dependence of the phonon frequencies?

We quantified the fluence-dependent amplitudes and relaxation times to explore the possible origin. Below $F_c$, the transient reflectivity can be fitted with a triexponential function; for above $F_c$, the transient reflectivity needs to be fitted with a quad-exponential function due to the appearance of the second rise peak at around 2 ps [Fig. 4(a)]. Figures 4(d) summarize the fluence dependence of amplitudes of the second components. For HF compounds, it has been suggested that, at low $T$, the amplitude $A$ and pump fluence $F$ are related by [41]

$$A \propto \sqrt{1+cF}-1 \tag{6}$$

where $c$ is a fitting parameter. At low fluence, -$A_2$ grows monotonically with increasing fluence. We found that Eq. (6) can account for the amplitudes at low fluence, as shown by the dashed green line in Fig. 4(d). We believe that, at low fluence, Kondo physics dominated the ultrafast dynamics, while at higher fluence, -$A_2$ deviates from the prediction of the RT model as the fluence continues to increase, with -$A_2$ even decreasing.

Figure 4(e) displays the fluence dependence of the relaxation times $\tau_2$, which differs from its temperature-dependent behavior [Figs. 2(a2) and 2(b2)]. Without photoinduced phase transition, amplitudes and relaxation times' fluence and temperature dependence are usually the same in heavy fermion materials [41, 56]. At low fluence, the hybridization measure $\tau_2$ decreases with increasing fluence, which can be well described by the RT model (dashed green line) [41, 56], indicating a gradual closing of the hybridization gap. At higher fluence, $\tau_2$ even increases, deviating from the predictions of the RT model. Note that the $\omega_1$ and $\omega_2$ frequencies are almost fluence independent, which means that the fluence-induced smearing of the gap could not be due to the heating effect mentioned above. Calculations by Pokharel *et al.* based on a simple steady-state thermal diffusion model [57] found that the average laser heating amounts of $YbB_{12}$ were only 1.1 degrees at a pump fluence similar to our experiments (0.5 mJ/cm$^2$). Thus, it is conceivable that the heating effect can be neglected even at the maximum excitation density used in our experiments. Therefore, we propose that the phonon frequencies anomaly is due to an intense pumping-induced rapid nonthermal phase transition, such as the photoinduced quenching of the magnetic order, accompanied by transient hybridization gap smearing. Further studies are needed to determine the nature of the possible photoinduced phase transitions.

In summary, we have investigated the ultrafast photoexcited quasiparticle dynamics in antiferromagnetic Kondo lattice compound $CeAuSb_2$ as a temperature and pump fluence function. Temperature-dependent quasiparticle relaxation reveals a hybridization gap of $\sim$ 4.5 meV, which was settled at $T^* \approx 100$ K. The anomaly of decay times, amplitudes, and phonon frequencies occurred at the characteristic temperatures $T^*$, $T'$, and $T_N/T_{nem}$, indicating that the Kondo effect and magnetic orders have essential roles in the relaxation process. We also observed a possible photoinduced nonthermal phase transition under high pumping fluence. We believe these findings provide critical information for a better understanding of heavy fermion physics and related fields.

## ACKNOWLEDGMENTS

This work was supported by the National Natural Science Foundation of China (Grants No. 12074436 and No. 11874264), the National Key Research and Development Program of China (Grant No. 2022YFA1604204), the Natural Science Foundation of Shanghai (Grant No. 17ZR1443300), and the Science and Technology Innovation Program of Hunan province (Grant No. 2022RC3068).

* Corresponding author: jqmeng@csu.edu.cn

# Supplemental Material:

## Coupling of optical phonons with Kondo effect and magnetic orders in antiferromagnetic Kondo lattice $CeAuSb_2$

Yin-Zou Zhao,[1] Qi-Yi Wu,[1] Chen Zhang,[1] Bo Chen,[1] Wei Xia,[2] Jiao-Jiao Song,[1] Ya-Hua Yuan,[1] Hao Liu,[1] Fan-Ying Wu,[1] Xue-Qing Ye,[1] Hong-Yi Zhang,[1] Han Huang,[1] Hai-Yun Liu,[3] Yu-Xia Duan,[1] Yan-Feng Guo,[2,4] Jun He,[1] and Jian-Qiao Meng [1,*]

[1]*School of Physics, Central South University, Changsha 410083, Hunan, China*
[2]*School of Physical Science and Technology, ShanghaiTech University, Shanghai 201210, China*
[3]*Beijing Academy of Quantum Information Sciences, Beijing 100085, China*
[4]*ShanghaiTech Laboratory for Topological Physics, ShanghaiTech University, Shanghai 201210, China*

**The supplemental materials to 'Coupling of optical phonons with Kondo effect and magnetic orders in antiferromagnetic Kondo lattice $CeAuSb_2$' contains 'Temperature dependence of resistivity ρ', 'Temperature-dependent of transient reflectivity at pump fluence of 30 μJ/cm$^2$', 'Temperature-dependent of transient reflectivity at pump fluence of 150 μJ/cm$^2$', 'Exponential decay fitting', 'Temperature dependence of phonon oscillations phase and damping rate', 'Raman spectrum of $CeAuSb_2$', and 'Calculated phonon dispersion and phonon modes'.**

## 1. Temperature dependence of resistivity $\rho$

Fig. S1 displays the temperature dependence of the in-plane resistivity $\rho$ of $CeAuSb_2$. The black arrow marks a sharp drop in the resistivity, indicating the antiferromagnetic transition temperature $T_N$ ~ 7 K. The gray arrow marks the broad shoulder around 13 K, discussed in the main text.

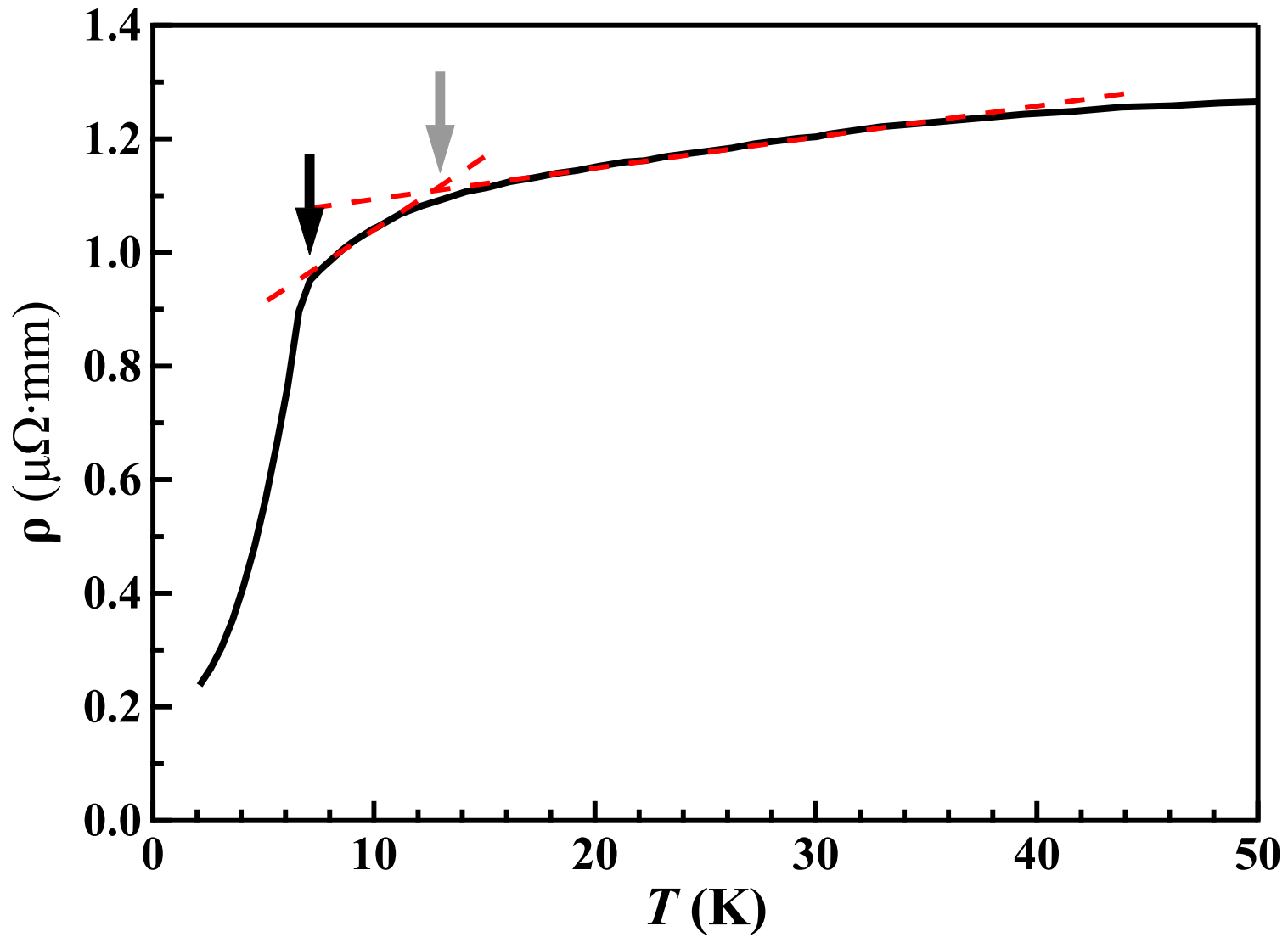

**Fig. S1** Temperature dependence of in-plane resistivity $\rho$ of $CeAuSb_2$ single crystal. The black and gray arrows mark a sharp drop and a broad shoulder, respectively. Red dashed lines are guides to the eyes.

## 2. Temperature-dependent of transient reflectivity at pump fluence of 30 μJ/cm$^2$

To highlight the initial peaks after ultrafast pumping, Fig. 1(a) was replotted in Fig. S2. Obviously, the intensity of the initial peak decreases as the temperature decreases. Even at the lowest temperature, the initial peak is still observable.

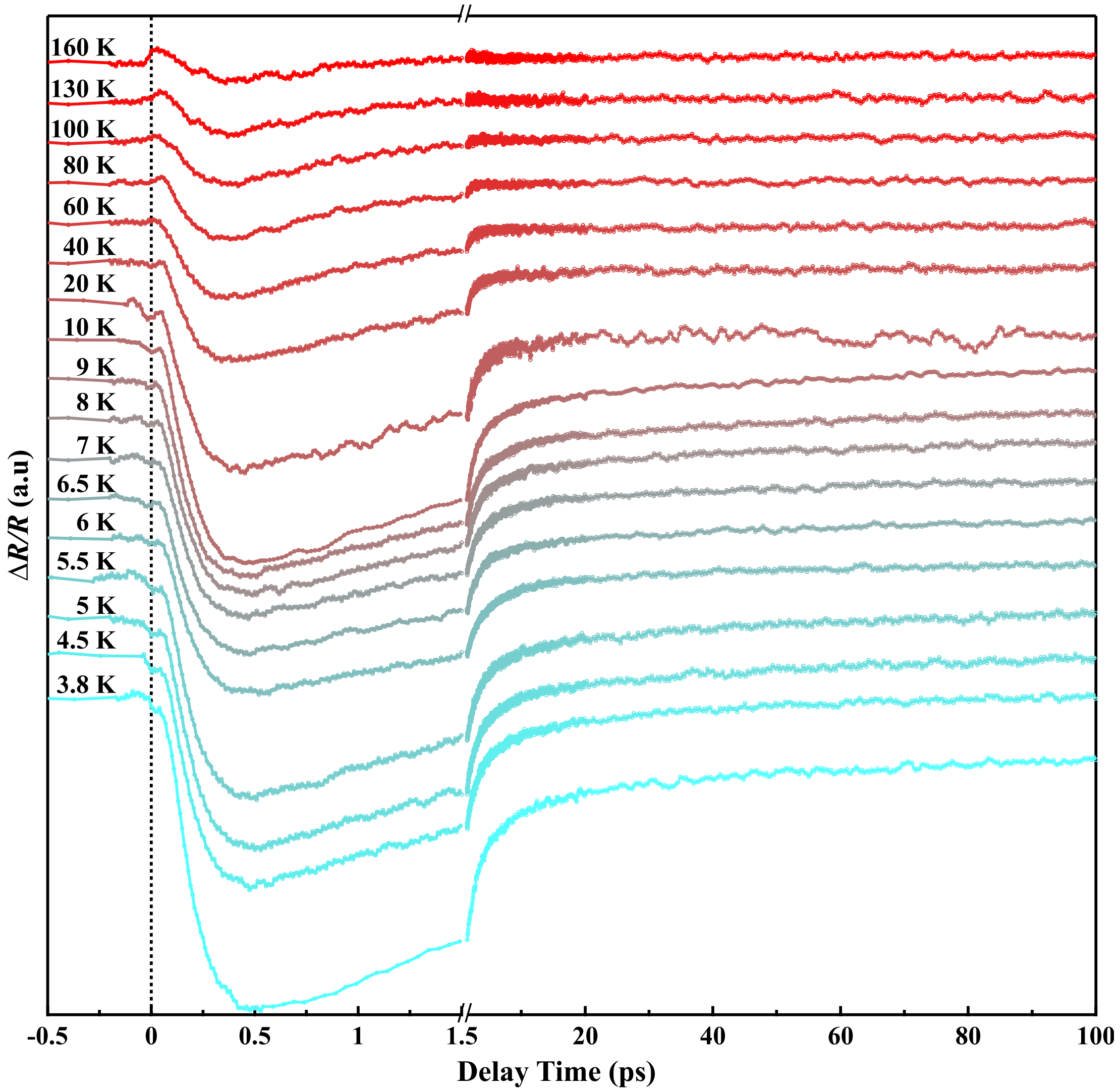

**Fig. S2** $\Delta R/R$ of $CeAuSb_2$ as a function of temperature at a pump fluence of 30 μJ/cm$^2$. Note the break in the *x* axis.

## 3. Temperature-dependent of transient reflectivity at pump fluence of 150 μJ/cm$^2$

Fig. S3 shows the measured *ΔR/R* results at a pump fluence of ~150 μJ/cm$^2$. The oscillation can be seen at all the measured temperatures. The intensity of the initial peak decreases as the temperature decreases. Even at the lowest temperature, the initial peak is still observable.

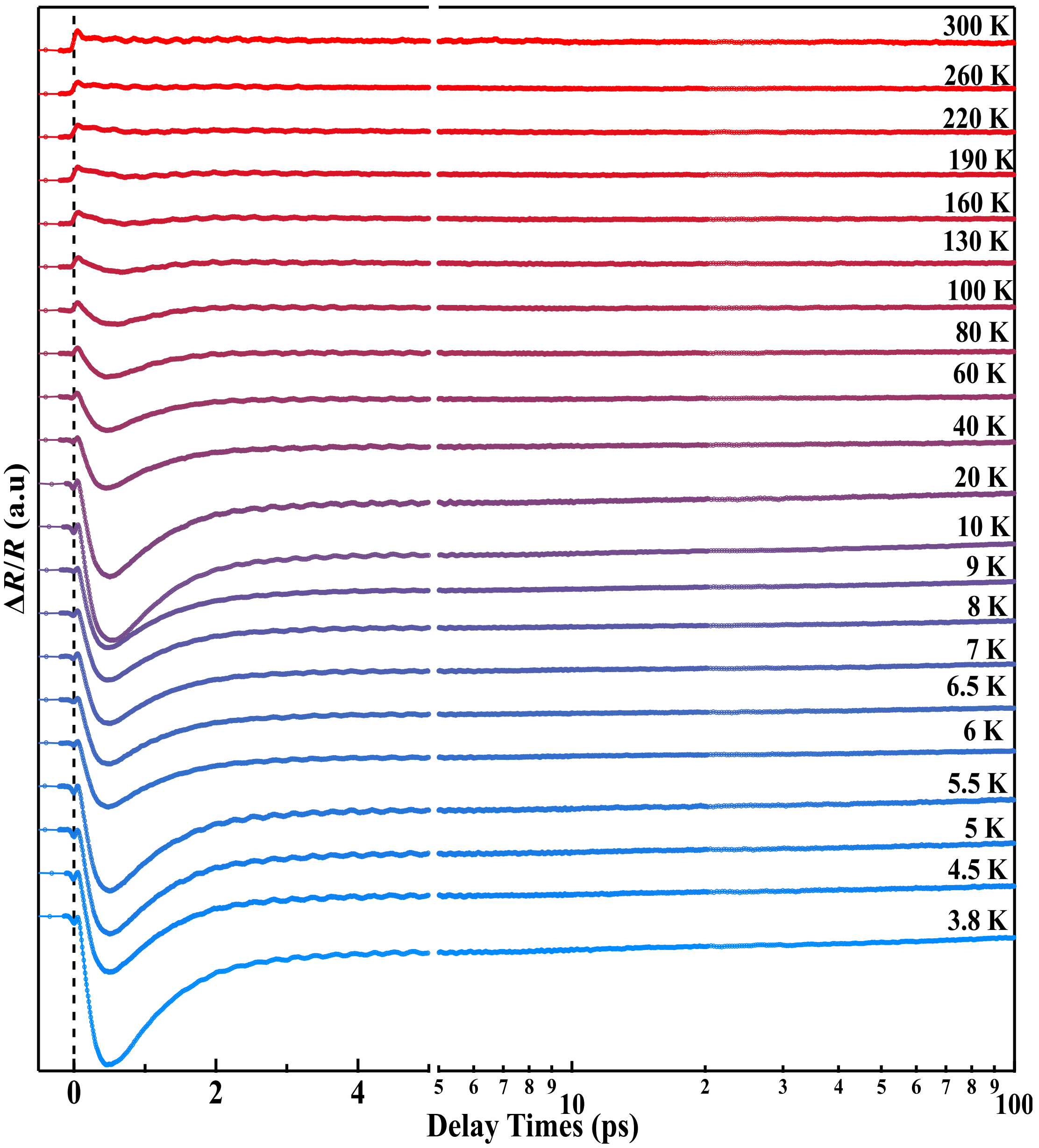

**Fig. S3** *ΔR*/*R* as a function of delay time over temperature range from 3.8 to 300 K at a pump fluence of 150 μJ/cm$^2$.

## 4. Exponential decay fitting

Fig. S4 shows the results of fitting the transient reflectivity at a pump fluence of 30 μJ/cm$^2$ with tri- or quad-exponential decay at selected temperatures [Eq. (1) in the main text]. It can be seen that the fit can be done well with tri-exponential at low temperatures (80 and 100 K). While at high temperatures (130 and 160 K), the hump caused by the apparent second rise resulted in the requirement to use a quad-exponential decay for the fit. Therefore, we used the quad-exponential function to analyze the data at higher $T$ and the tri-exponential function to analyze the data at $T \leqslant 100$ K ($A_3$ was set to zero).

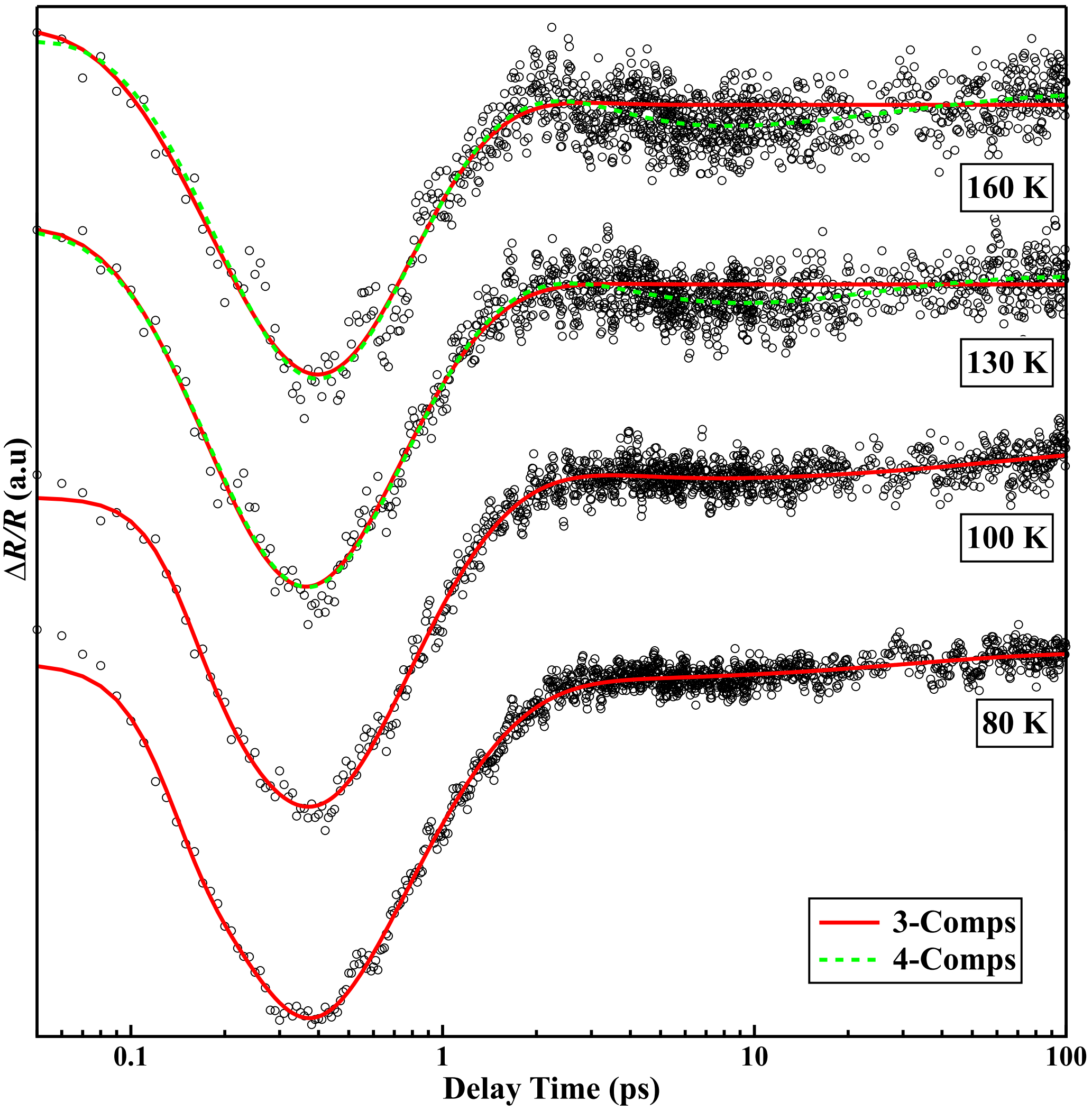

**Fig. S4** The results of the fitting of four selected temperatures data sets using the tri- or quad-exponential decay function. Red solid and green dashed lines represent the fitting curves with tri- and quad-exponential function, respectively.

Fig. S5 shows the results of fitting the transient reflectivity at 3.8 K with tri- or quad-exponential decay at selected pump fluence. It can be seen that the fit can be done well with tri-exponential at lower pump fluence (150 and 180 μJ/cm$^2$). While at higher pump fluence (240 and 300 μJ/cm$^2$), the hump caused by the apparent second rise resulted in the requirement to use a quad-exponential decay for the fit. Therefore, we used the quad-exponential function to analyze the data at higher pump fluence, and we used the tri-exponential function to analyze the data at $F \leq 180$ μJ/cm$^2$ ($A_3$ was set to zero).

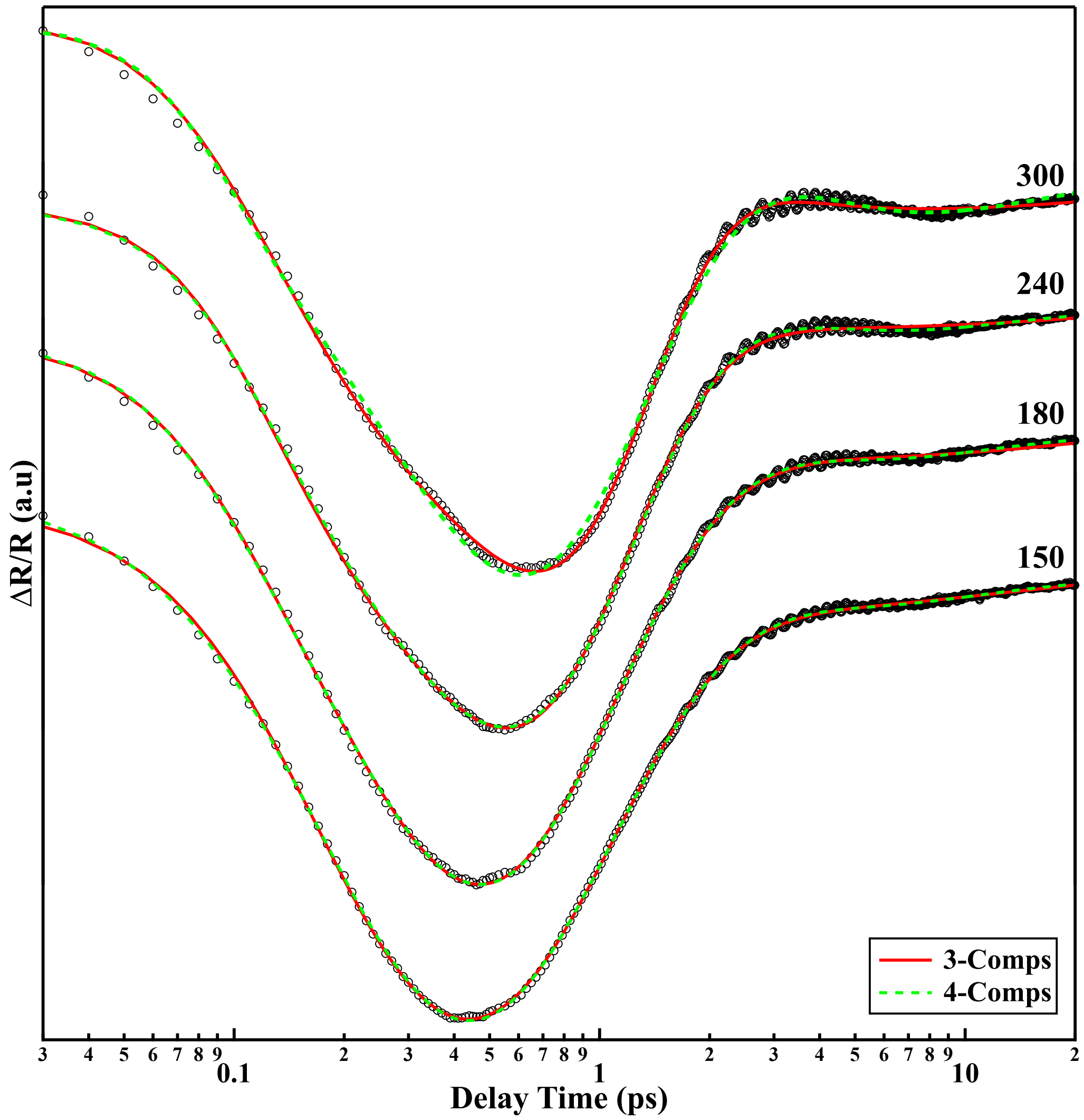

**Fig. S5** The results of the fitting of four pump fluence data sets using the tri- or quad-exponential decay function. Red solid and green dashed lines represent the fitting curves with tri- and quad-exponential function, respectively.

## 5. Temperature dependence of phonon oscillations phase and damping rate

Fig. S6 shows the derided phase ($\varphi_1$ and $\varphi_2$) and damping rate ($\Gamma_1$ and $\Gamma_2$) of phonon oscillations [$B_{1g}$ ($\omega_1$) and $E_g$ ($\omega_2$)] as a function of temperature at a pump fluence of 150 μJ/cm$^2$. Considering that the phonon frequencies of $\omega_1$ and $\omega_2$ are similar, fitting $(\Delta R/R)_{OSC}$ would introduce a large error bar. Therefore, as shown in Fig. S6(a), we consider the phonon oscillation phase to be a constant over the entire temperature measurement range, with $\varphi_1$ and $\varphi_2$ being 0 and π, respectively. This procedure is valid because $\varphi$ must be either 0 or π, corresponding to the amplitude's positive or negative sign, respectively. For the damping rates shown in Fig. S(b), there appear to be some changes occurring around $T^*$(≈ 100 K) and $T'$ (≈ 13 K). However, since the error bars here are significant, we do not want to say too much.

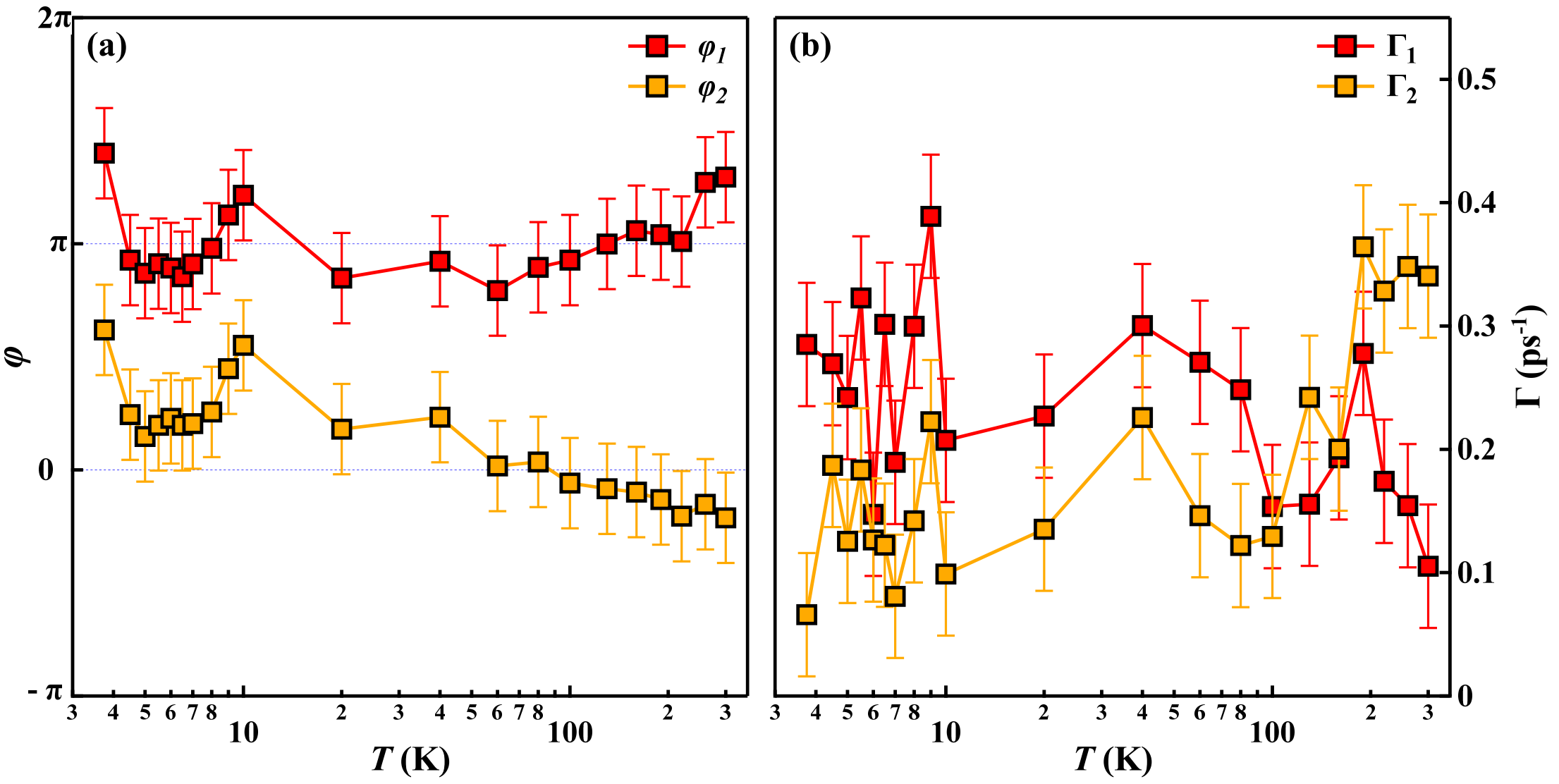

**Fig. S6** Temperature dependence of the $B_{1g}$ (red squares) and $E_g$ (yellow squares) phonon oscillations phase (a) and damping rate (b) extracted from the $(\Delta R/R)_{OSC}$ data.

## 6. Raman spectra of $CeAuSb_2$

Fig. S7 (a) displays the Raman spectra of $CeAuSb_2$ single crystal obtained in the backscattering configuration under 633-nm laser excitation. Five phonon modes are observed in a range of 20 to 220 $cm^{-1}$. At 80 K, the phonon modes at ~42.2, ~66.4, ~116.1, ~121.4, and 151.9 $cm^{-1}$ can be assigned to $E_g$, $B_{1g}$, $E_g$, $B_{1g}$, and $E_g$ modes, respectively, based on the comparison of Raman spectra with the calculated phonon dispersion. Frequencies were estimated by fitting the Raman spectra peak. Fig. S7 (b) shows the detailed temperature dependence of position for phonons.

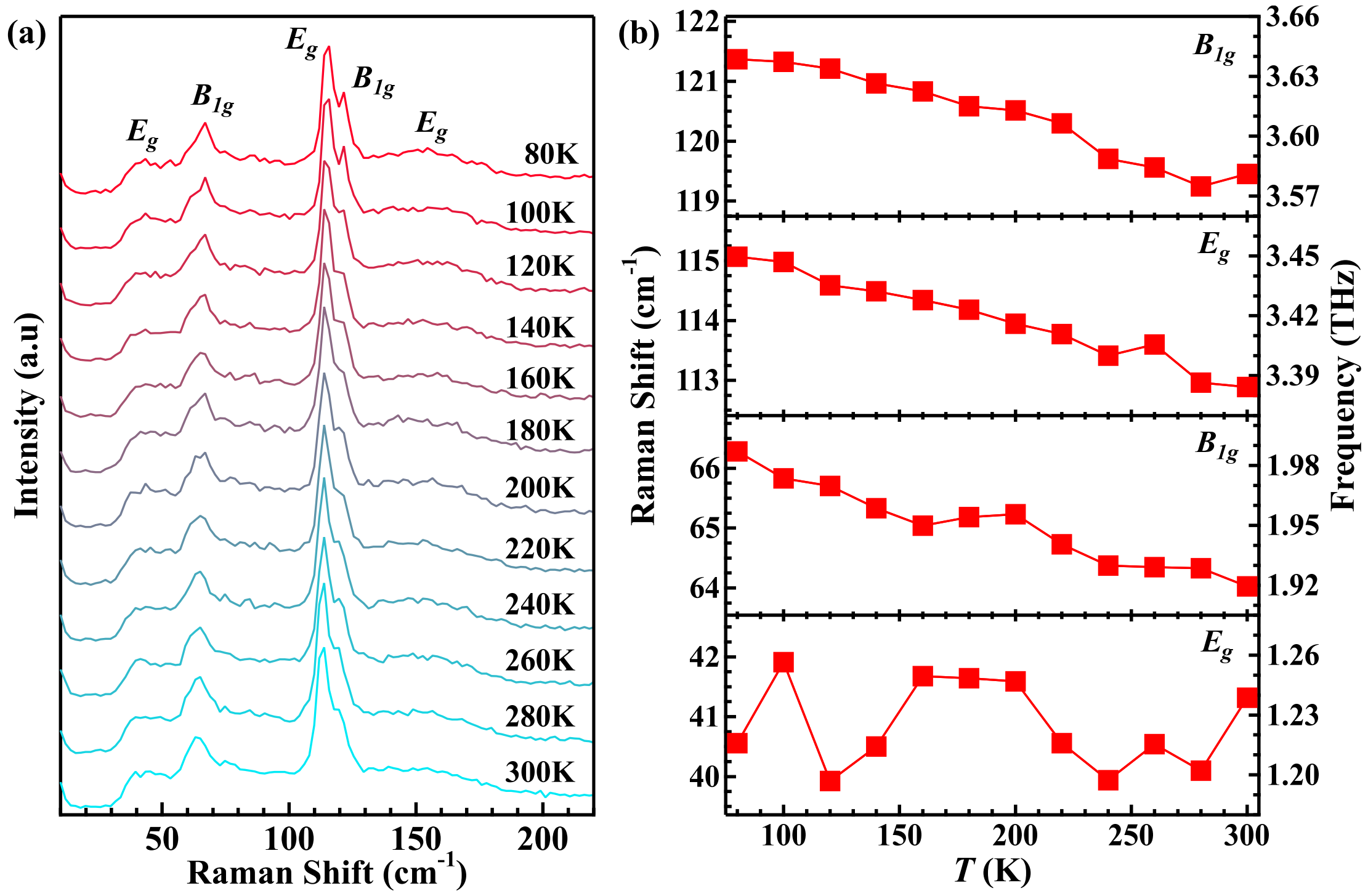

**Fig. S7** Raman shift of phonon modes of $CeAuSb_2$. (a) Raman spectra at marked temperatures. (b) Temperature-dependence of Raman shift of phonon modes.

## 7. Calculated phonon dispersion and phonon modes

The calculated phonon dispersion curves along several high symmetry directions for the $CeAuSb_2$ are shown in Fig. S8. The calculated phonon frequencies and the experimental phonon modes at 80 K are given in Table I. The layered $CeAuSb_2$ has eight Raman-active modes at Γ point: $2A_{1g} + 2B_{1g} + 4E_g$.

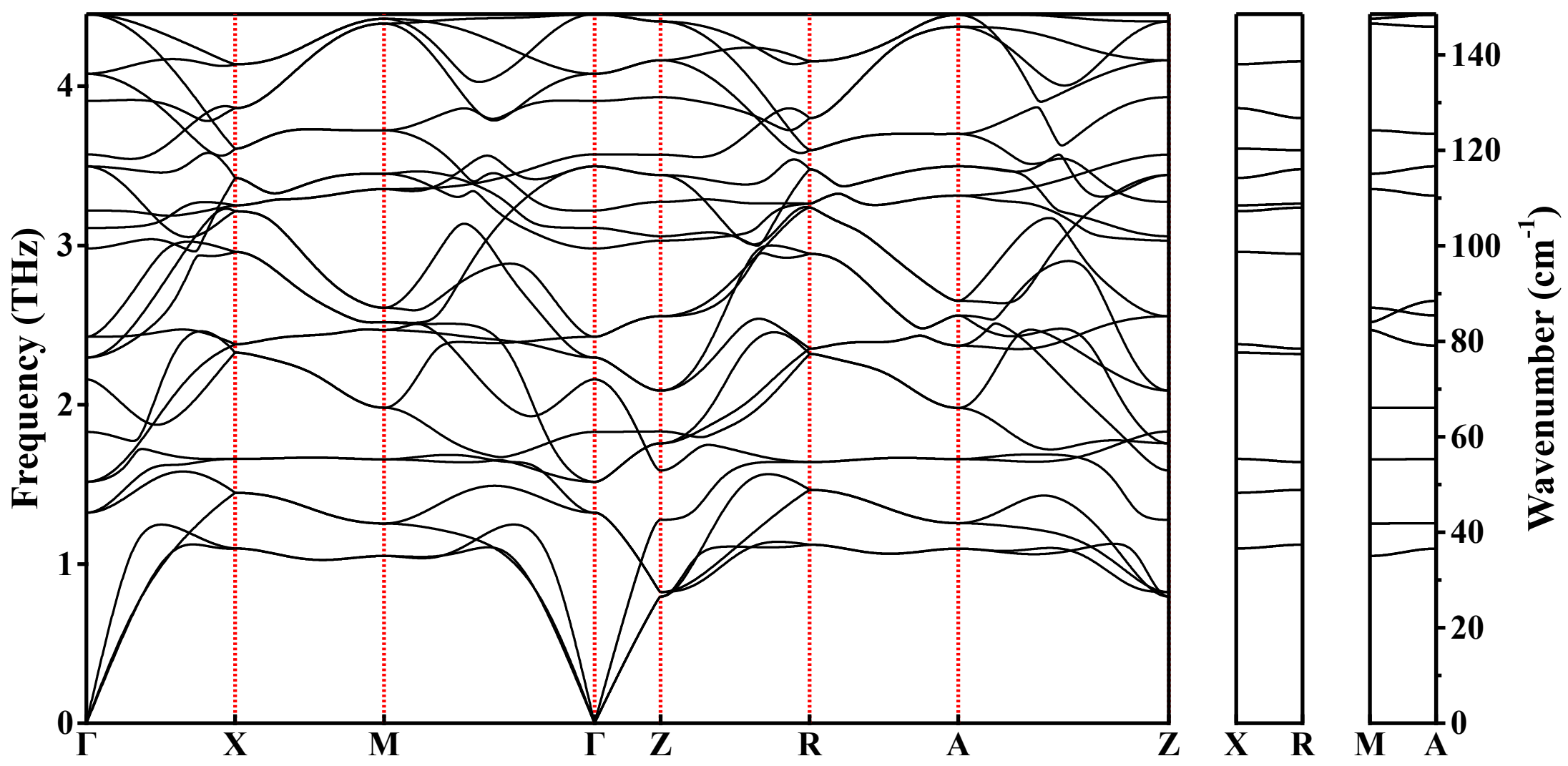

**Fig. S8** Phonon dispersion calculated by density function perturbation theory.

**TABLE I.** Comparison of the calculated and experimental of phonon modes.

| Sym. | Cal. (THz) | Cal. ($cm^{-1}$) | Exp. ($cm^{-1}$) | Main atom displacements[a,b] |
|---|---|---|---|---|
| $E_g$ | 1.32 | 44.0 | 42.2 | Au($x$), $Sb^1$($-x$) |
| $E_u$ | 1.52 | 50.7 | | Au($x$), $Sb^2$($-x$) |
| $B_{1g}$ | 1.83 | 61.0 | 66.4 | Au($z$) |
| $A_{2u}$ | 2.16 | 72.0 | | Au($z$), $Sb^1$($z$), $Sb^2$($-z$) |
| $E_u$ | 2.30 | 76.7 | | Sb1($-x$), $Sb^2$($x$) |
| $E_g$ | 2.43 | 81.1 | | Ce($x$), Au($x,-x$) |
| $A_{1g}$ | 2.98 | 99.4 | | Ce($z$), $Sb^1$($-z$) |
| $A_{2u}$ | 3.11 | 103.7 | | Au($z$), $Sb^1$($-z$) |
| $A_{1g}$ | 3.22 | 107.4 | | Ce($z,-z$), $Sb^1$($z,-z$) |
| $E_g$ | 3.50 | 116.7 | 116.1 | $Sb^2$($y,-y$) |
| $B_{1g}$ | 3.57 | 119.1 | 121.4 | $Sb^2$($z,-z$) |
| $A_{2u}$ | 3.91 | 130.4 | | Ce($z$), $Sb^2$($-z$) |
| $E_u$ | 4.08 | 136.1 | | Ce($x$), $Sb^1$($-x$) |
| $E_g$ | 4.45 | 148.4 | 151.9 | Ce($x$), $Sb^1$($-x$) |

[a] $Sb^1$ in the Ce-Sb layer

[b] $Sb^2$ out of the Ce-Sb layer